# Situational Awareness Enhanced through Social Media Analytics: A Survey of First Responders

Luke S. Snyder
Visual Analytics for Command, Control, and Interoperability Environments (VACCINE) Center
Purdue University
West Lafayette, Indiana, USA
snyde238@purdue.edu

Morteza Karimzadeh
Department of Geography,
University of Colorado Boulder
Boulder, Colorado, USA
karimzadeh@colorado.edu

Christina Stober
Visual Analytics for Command, Control, and Interoperability Environments (VACCINE) Center
Purdue University
West Lafayette, Indiana, USA
cstober@purdue.edu

David S. Ebert
Visual Analytics for Command, Control, and Interoperability Environments (VACCINE) Center
Purdue University
West Lafayette, Indiana, USA
ebertd@ecn.purdue.edu

*Abstract*—Social media data has been increasingly used to facilitate situational awareness during events and emergencies such as natural disasters. While researchers have investigated several methods to summarize, visualize or mine the data for analysis, first responders have not been able to fully leverage research advancements largely due to the gap between academic research and deployed, functional systems. In this paper, we explore the opportunities and barriers for the effective use of social media data from first responders' perspective. We present the summary of several detailed interviews with first responders on their use of social media for situational awareness. We further assess the impact of SMART—a social media visual analytics system—on first responder operations.

*Keywords—situational awareness, social media, survey*

## I. INTRODUCTION

There has been considerable research on the potential and use of social media data for situational awareness, such as disaster response, event monitoring, or public sentiment analysis [1–5]. In particular, researchers have provided visual analytics approaches combining visualization and data analysis [2–7], geoparsing for crisis-mapping [7, 8], cloud-based mobile services for on-the-ground information [14], and data mining and natural language processing techniques such as clustering and named entity recognition [15].

The majority of this research has focused on devising computational and/or visual methods for making sense of the vast amount of streaming social media data. However, the first responder community has not been able to fully leverage the state-of-the-art achievements made possible through such methods, mostly because (a) the advancements are not made available as user-friendly, operational and scalable systems, and (b) there is a gap between academic research (which strives to push the boundaries of science) and real-world requirements (which have different interests and may be constrained by various social, political, and ethical issues). In this paper, we build on our research on the development of the Social Media Analytics and Reporting Toolkit (SMART) [6, 9]—which has been used by many first responders—to explore the opportunities and obstacles in the use of social media data for real-time situational awareness. We leveraged our user community and conducted interviews with first responders tasked with monitoring or using social media data for situational awareness (e.g. to report on an unfolding situation, coordinate action on the ground, and pass on information to relevant entities) to inform research and development in this field. The results of our interviews highlight the advantages and shortcomings of using social media data analytics for situational awareness in real-world operational settings.

In what follows, we summarize the responses we received from interviewees in three major categories: (1) The use of social media platforms in first responder operations; (2) The impact of SMART social media analytics on first responder operations; and (3) Detailed feedback on the usefulness of SMART's features. The interviewees' responses shed light on future directions for successful research, development and adoption of social media analytics for situational awareness.

## II. METHODS

We interviewed 13 first responders who had been introduced to and/or worked with SMART for situational awareness in emergency operation centers or command centers during the 2016-2019 period. All interviewees were active members of either law enforcement (8) or the United States Coast Guard (USCG) (5). Interviews were conducted over video calls or in-person, based on the location of interviewees. Per Institutional Review Board policies to protect human subjects, all survey responses have been anonymized to protect the interviewees' privacy by removing personally identifiable information (PII). As a result, some specifics from the responses have been removed from this paper.

## III. OVERVIEW OF SMART

SMART [6, 9] was developed at Purdue University through an iterative, user-centered design by collecting feedback from the first responder community to ensure its usefulness. SMART provides a user-friendly, graphical web-based user interface, hiding the complexity of advanced algorithms filtering and summarizing social media data, which allows users to interactively explore, filter, and visualize trends and anomalies in real-time geo-tagged Twitter data. SMART also provides topic aggregation, natural language processing, and probabilistic event detection to analyze and refine the high volume of data.

SMART has been successfully deployed by first responders during several major events in the United States, including holiday celebrations, political conventions, the 2017 Presidential Inauguration, and multiple Presidential State of the Union events. Using SMART, first responders have identified actionable information such as road closures for further investigations. Additionally, the USCG utilized SMART during the four hurricanes that made landfall in the U.S. in 2017, allowing them to respond more efficiently to developing situations.

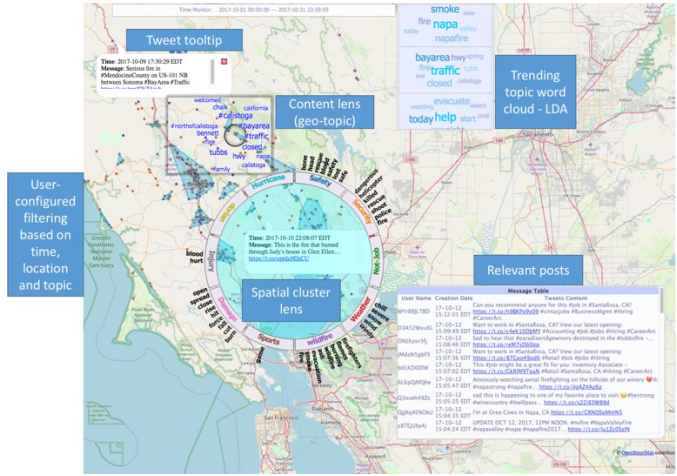

Fig. 1. SMART in a wildfire use-case. User-configured classifiers for keyword filtering, content lens for word cloud (topic) visualization in specific geographical locations, cluster lens for classifier keyword detection within tweet clusters, heat mapping to view the spatial distribution of tweets, temporal view (time bar graph) to visualize the time-evolution of tweet content, and the world cloud (LDA) topic modeling view to identify trending topics and associated keywords.

## IV. SURVEY RESULTS: USE OF SOCIAL MEDIA AND CROWDSOURCING BY FIRST RESPONDERS

In this section, we review the results of our survey on the use of social media for situational awareness, using the platforms directly or tools other than SMART. In the next major section, we will focus on SMART's usefulness for this purpose.

### A. Purpose of/Reasons for Monitoring Social Media

Most of the interviewees stated that they monitor social media (in general, or using SMART, specifically) during large events (such as street festivals, holiday events, or inclement-weather situations) to improve their situational awareness. Social media monitoring provides real-time information about developing situations, allows first responders to observe the overall atmosphere of the event, and may provide updates more quickly than official reports or calls from ordinary citizens; therefore, several interviewees found it useful for developing situational awareness more rapidly and efficiently than other methods such as calls from the public or patrols.

According to our interviews, social media helps identify warnings or possible action-times, depending on the category. Safety/health issues can trigger immediate actions if reported on social media, while law enforcement concerns prompt further monitoring or corroboration in situations involving crimes.

Large events and natural disasters require first responders to deal with rapidly changing situations and huge amounts of data. Social media, if filtered based on first responder needs, can provide information that might not otherwise reach officials in a timely manner. Some of our respondents reported that they monitor specific threats or concerns based on previous events; others monitor social media for a general overview of situations, good or bad, that might require a response from their agency or a referral of information to other related agencies (e.g. health and safety, law enforcement, fire department).

### B. Information Handling Process

In this section of the survey, we discussed interviewees' typical information-gathering processes, including all social media monitoring as well as information from non-social media sources, and the ways those processes and information sources interacted. Some replies incorporated SMART monitoring as part of analyzing their overall processes.

*1) Informational Channels: Social Media and Other Sources*

Only three of our interviewees used software platforms other than SMART specifically intended to monitor social media. Tweetdeck [16] and SocialBearing [17] were used for monitoring and Hootsuite [18] was used for posting to multiple departmental social media accounts simultaneously.

Some interviewees searched social media platforms, e.g. Facebook or Twitter, directly; others relied on newsfeeds, posts or messages directly contacting responders, call stations set up to direct information to the relevant command centers, and official information sources such as the National Weather Service, utility company reports, and notices from other agencies.

*2) Corroboration or Conflict between Social Media and Other Sources*

Interviewees reported that official news sources often corroborated information gleaned from social media (such as street closures or power outages reported on social media first). Similarly, assumptions about the overall event proceedings/mood based on official sources were often confirmed on social media that reflected people's real-time experience (for example, at an event with low attendance due to high heat and humidity, "weather" was one of the largest categories mentioned in social media posts).

Five interviewees reported that information gleaned from social media was often more up-to-date than the official sources, which sometimes had a lag in reporting. For example, social media posts might indicate blocked roads before calls to emergency operation centers or before first responders explored the area; or traffic issues that might have gone unreported are often mentioned on social media. However, according to interviewees, social media posts tend to be much less specific than the official news sources and usually require follow-up or corroboration in order to be useful for direct action.

Some interviewees had concerns about the veracity of social media information; and therefore, implemented external processes to cross-check against officially-derived information, such as sending personnel directly to the scene or contacting related agencies for status updates.

*C. Actions*

In this section, we review interviewees' actions taken based on the information-gathering processes detailed above. Some specific examples given in the surveys have been genericized to protect interviewees' privacy based on IRB guidelines.

*1) Criteria for Dispatch/Action*

Almost all of the interviewees stated that they did not take direct action based solely on social media, either for policy reasons (law enforcement agencies cannot obtain warrants based on social media posts) or due to concerns about the accuracy of information gained from social media. However, several interviewees had implemented verification processes including sending personnel to monitor affected areas, where they could also respond to developing situations if necessary. Social media monitoring was generally used to either prompt further investigation or confirm/update information from other sources.

In some cases, posts of concern were passed on to relevant law enforcement or safety departments. For example, during severe weather events, social media posts about power outages were passed on to the power utility company (after appropriate verification) so that the issue could be remedied quickly.

*2) Successful Action based on Information Gleaned from Social Media*

While most interviewees did not take direct action based solely on social media, some posts did serve as prompts for further investigation. During severe weather events, social media posts might indicate downed trees or cars stranded in flooded areas; emergency operation centers could send out emergency services to verify and assist, or contact the relevant agency. Urgent medical issues (e.g. heatstrokes) have also been flagged through social media, allowing first responders to send medical personnel to the area to assist.

*3) Erroneous Action based on Information Gleaned from Social Media*

None of our interviewees reported any erroneous action based on social media information. Many of them cited the policy limits on acting based on social media as the reason, meaning that only information whose veracity was confirmed was used for direct action.

*D. General Reflections*

*1) Major Advantages and Challenges of Monitoring Social Media*

When asked about advantages, many interviewees mentioned that social media provides a fuller picture of the situation, a sense of how the community is feeling and responding, as well as immediate concerns that might take time to be reported through official channels. Some interviewees also found it useful for investigation after the fact, and for extra information that could supplement official reports or records.

When it came to challenges, several interviewees commented on the sheer amount of information coming through social media, and the need for specially-trained responders who knew how to use the platforms, social media mining tools, and specific flags to look for on social media. Lack of manpower was consistently cited as an issue, as monitoring social media requires having at least one person dedicated to the job and many agencies are already stretched thin. The possibility of missing things in the flood of data is a big concern, as is the fact that tools and social media platforms often limit the amount of information that can be accessed in the first place. Hoax/fake posts and certain types of slang can cause errors in filtering; for example, tools might flag the phrase "shoot on over for a free pizza" because of the word "shoot", while a post using an innocuous slang term for an illegal drug might be bypassed entirely due to not including any of the predetermined keywords.

*2) Missing/Desired Information*

Several interviewees mentioned Facebook specifically as a site their agencies are interested in monitoring, since it is a widely used social media platform for some regions and population brackets. Due to Facebook's terms of service, however, monitoring programs cannot access Facebook posts, and legal or departmental restrictions may prevent agencies from directly monitoring the platform.

Prohibitions in social media providers' terms of service often limit official/government access and monitoring, and the majority of social media platforms have privacy options that can prevent outside agencies from viewing specific posts or profiles. This allows privacy for citizens but may prevent first responders from quickly reacting to a situation due to not having enough timely information.

Geographic information is not always reliable or particularly specific; for example, individuals might post on social media about a power outage caused by a storm, but the posts will not provide disaster response teams with information about how many houses are affected or what precisely caused the outage (a downed power line, an issue at the power plant, an issue at the house itself, etc.). Several interviewees mentioned a desire for better location information, which again, is a challenge due to privacy concerns.

Many of our interviewees also mentioned the ability to monitor live street traffic as an interest and mentioned that the ability to link systems together so that tools monitoring Twitter, Facebook, live traffic, official weather alerts, etc. could all be viewed in one central location would be very useful.

*3) Policy and Regulation Concerns*

Many of our interviewees mentioned the lack of legal and policy clarity and inconsistency around first responder access to/use of social media posts as a concern. Interviewees described situations in which the legal and policy compliance permissibility of accessing and acting on certain information is unclear, which may result in accessing information they legally should not, or missing information that they could have used. Some interviewees also mentioned a need for individual departmental policies to supplement the legal and political contexts. Several interviewees also raised ethical concerns surrounding privacy; especially in a law enforcement context, several interviewees had concerns about maintaining citizen privacy while still accessing necessary information.

## V. SURVEY RESULTS: IMPACT OF SMART ON USCG AND FIRST RESPONDER OPERATIONS

In this section, we focus on the usefulness of SMART for situational awareness. At the time of this writing, SMART supports 376 users who may sporadically use the system. Throughout SMART's development and lifecycle, users have consistently attested to its value and effectiveness and continued to utilize it. In this section, we provide an overview of when and how users have applied SMART, and their overall opinion of the system.

### A. Events with SMART Use

SMART has been used in many events, but most notably the following: Presidential Inauguration 2017, State of the Union 2017/2018/2019, Baltimore Fleet Week 2016, Coast Guard maritime events, Cincinnati's annual Riverfest, Thunder over Louisville's airshows and fireworks, 4th of July celebrations and other holiday events, Inclement weather (e.g. hurricane) events and multiple college football games.

### B. SMART Utilization

#### 1) SMART's Role in Event Monitoring

Interviewees who had used SMART incorporated it into their event command centers or emergency operations centers. Dedicated staff were tasked with situational awareness and coordination, sometimes monitoring SMART only, and sometimes monitoring other social media or newsfeeds. For large events, some recurring/long-term users used advanced monitoring leading up to the event, with 6-hour summaries and email alerts. Several newly introduced users also expressed interest in this feature, though they had not had the opportunity to use it as of the time we conducted our interviews. Most of the interviewees focused on smaller-scale monitoring (local events or games) and designated an individual to monitor and report anything of concern identified through the use of SMART.

#### 2) SMART's Relationship to Other Tools

None of the surveyed users directly linked SMART to other tools. Some of them monitored it alongside other tools (for example, the national weather service, and a departmental Facebook account) but did not connect them in any way aside from comparing the occasional instance of overlapping information.

### C. Process and Actions

#### 1) SMART and Other Information Sources

Most users reported no significant overlap between SMART and other information sources. In a few cases, it corroborated information (developing bad weather and medical issues) that had been received through phone calls or news feeds. Sometimes SMART provided complimentary, general information, and official reports and phone calls more specific information. No users reported instances where information gathered from SMART conflicted with other information sources.

#### 2) Actions Initiated by SMART

Most users did not take specific actions solely based on information gleaned from SMART, though SMART consistently provided timely information for further investigation when an incident did arise. Actions were typically based on a broader situation with information obtained or verified using other sources. SMART sometimes prompted responders to look further into a specific situation, but more often, users relied on it to expedite decision-making by improving awareness of the overall situation. Some first responders passed on tweets of concern to relevant agencies (for example, if SMART flagged a tweet about flash floods during a severe weather event, first responders might contact roadwork companies, or have a PR team send out safety messages to warn people in the area not to drive).

#### 3) Successful Use of SMART

Several users reported that SMART expedited their decision-making by up to 15-20 minutes during large events and severe weather, since they were able to use it to monitor developing situations with real-time data. In several cases, credible concerns identified through SMART prompted further research that might not have been performed in the absence of social media information.

#### 4) Erroneous Use of SMART

All of our interviewees stated that no (resulting) erroneous action had been taken based on SMART.

### D. General Reflections

#### 1) Pros and Cons of Using SMART

Several users listed the customized search keywords/indicators and the real-time analytics as major benefits of using SMART. First responders covering areas where Twitter (and Instagram linked to Twitter) are among major social media platforms mentioned that SMART's focus on these social media platforms made it more useful than other monitoring tools. Long-term users of SMART (who have used the system since as early as 2012) mentioned the consistent Purdue University research and development support and the ability to have one or two people monitoring it as benefits, as well as the ability to develop a consistent pattern for using the system and incorporating it into their event planning.

Several interviewees mentioned that the system does require training and is not particularly intuitive. However, all but three repeat users of the system did not feel the need for re-training and all users were able to quickly re-familiarize themselves with the system again even if it had been months since the last time they had used SMART.

The SMART filters primarily use user-specified keywords to narrow down the stream of tweets to the topical relevant ones. Keywords, however, may cause some contextual issues; slang, such as a tweet using "bomb" to mean "awesome", may cause false positives, or synonyms such as "slaughter" rather than "kill", not be caught by the classifiers. Based on user feedback, while these interviews were in progress, we extended SMART with interactive machine learning-based classifiers to leverage context beyond just the syntax of the words and include the meaning (semantic) of the keywords in the filtering process, which we believe will significantly improve this issue in future uses of the system [6].

Two interviewees mentioned location accuracy issues with tweet geotagging, as a result of approximate geotagging done by the user (e.g., when a user geotags their tweets with the State or City they are in, instead of geographic coordinates or postal address). The lack of Facebook monitoring (which is prohibited by Facebook terms of service and lack of public API, and not a SMART limitation) was also referenced by several users, who saw value in harvesting Facebook posts for situational awareness. Furthermore, users had concerns about the small percentage of tweets that are publicly available through the Twitter Streaming API (which SMART collects) compared to the entire Twitter firehose, as well as the minority of tweets being geotagged (usually between 1% and 2% of the entire tweets, up to 10% during crisis times when people intentionally turn on their geotags).

*2) SMART Effectiveness*

All interviewees described SMART as effective for their goals, with features that were not actively used or required to execute the department's objectives being described as adequate (e.g., LDA topic modeling view and geofencing). Several mentioned that further training on some of the features such as the LDA topic modeling view and heatmap would likely enhance their use of the system, since most used it primarily for monitoring keywords and/or tweets in a specific geographic area rather than manually reviewing each individual tweet.

## VI. FEEDBACK ON SMART FEATURES

We asked our interviewees about their perception of the usefulness of SMART's individual features, as well as their frequency of use of each tool.

### A. Features Ranked by Frequency of Use

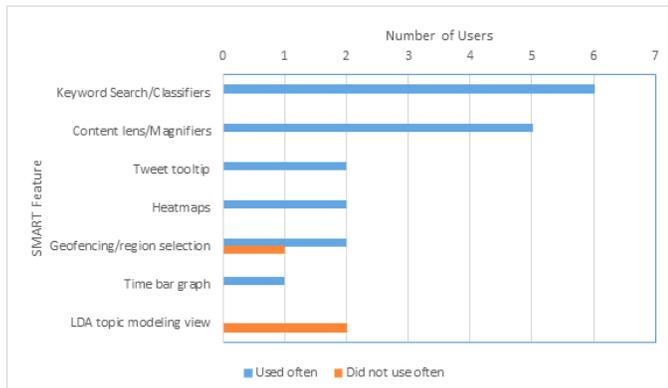

Fig. 2. SMART features ranked by the number of interviewees who frequently used them.

We used the feature names (e.g. content lens, heatmap) to ask our interviewees about their frequency of use and ranking. The interviewees who took the survey as part of a first-time demonstration were not able to provide a ranking due to having very little experience with the tool; and of those who did, many were unfamiliar with the official names for each tool. However, they were usually able to describe each feature even without knowing the official names, though in some cases we provided descriptions to resolve the ambiguity. Of the nine users who provided rankings, seven named the classifiers as a tool they used frequently, and five listed the content lens as well. The tweet tooltip was mentioned by only two users, but both described it as the tool they relied on most with other tools being secondary or less-frequently used. Only one user mentioned the temporal view (time bar graph), but used it as often as the classifiers and content lens. Two users mentioned the geofencing tool, but only one used it regularly. Two interviewees mentioned the LDA topic modeling view (i.e., word cloud) as a tool they did not use frequently (since the map already provides that functionality).

### B. Features Ranked by Usefulness

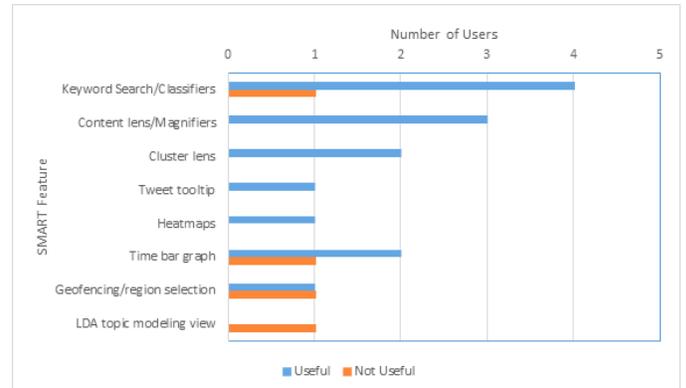

Fig. 3. SMART features ranked by the number of users who found them useful.

Again, the most useful feature was the keyword search/classifier, though one interviewee described it as "fairly effective" and would have considered it more useful if the graphical user interface were more intuitive. Three interviewees also found the content lens extremely useful. While no interviewee mentioned the cluster lens in response to the frequency of use, two mentioned it as useful when they did access it. The temporal view and region selection had similar results to the frequency of use, with some interviewees liking them and others finding them irrelevant or inaccurate. Only one interviewee mentioned the LDA topic modeling view (word cloud) and ranked it very low on the scale of usefulness.

## VII. CONCLUSIONS

In this paper, we provided an in-depth summary and discussion of key points from our interviews with several first responders tasked with situational awareness. Specifically, based on survey feedback, social media tools such as SMART play an important role in providing information that can be used to better assess and monitor events in a timely manner that might not be possible with major news sources. Interviewees found that social media often provided information up to 15-20 minutes before major news sources. Additionally, although social media was not used for direct action without corroboration, it did prompt further investigation for faster response.

Feedback was positive for SMART, with interviewees remarking on its effective visualizations in locating important information. Further, SMART has been successfully used in several national events, such as the 2017 Presidential Inauguration and hurricane events to more effectively facilitate situational awareness.

Interviewees, however, noted several limitations with SMART and social media in general. In particular, the majority of social media data is either not geotagged, making it difficult to determine the context of the message, or is unavailable due to privacy settings or lack of clarity on policies. Further, due to the vast amount of social media data and textual limitations such as misspellings or slang, gleaning relevant information is difficult.

Based on collective survey feedback, we recommend social media analytics systems focus on (1) improving the identification of relevant social media messages, (2) increasing the amount of streaming data by utilizing well-known social media APIs such as Twitter and computationally enriching the information without explicit geo-information, and (3) integrating information such as traffic data, weather alerts, and news updates in order to enhance situational awareness and provide increased context for end users.

For successful transition and independent utilization of social media analytics toolkits, the following additional steps would be beneficial: (1) Keeping the user manuals up to date through centralized hosting of online material (reflecting the latest version); (2) creating short training videos for introduction of each feature and for several use case scenarios; and (3) creating Standard Operating Procedures (SOP) material and documentation of the process to get approval for use of toolkits for different situations.

ACKNOWLEDGMENTS

This material is based upon work funded by the U.S. Department of Homeland Security (DHS) under Award No. 2009-ST-061-CI0003, DHS Cooperative Agreement No. 2014-ST-061-ML0001, and DHS Science and Technology Directorate Award No. 70RSAT18CB0000004.

APPENDIX

## A. SMART Features

In this appendix, we review SMART's features that first responders were surveyed on (Section VI of the main paper).

### 1) Keyword-based Filtering
#### a) Keyword Search

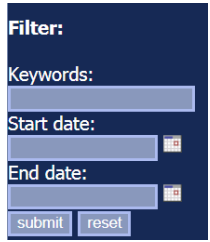

SMART users can enter multiple keywords of interest into the keyword filter box. Social media posts containing one or more user-input keywords are filtered and displayed in various views.

#### b) Classifiers

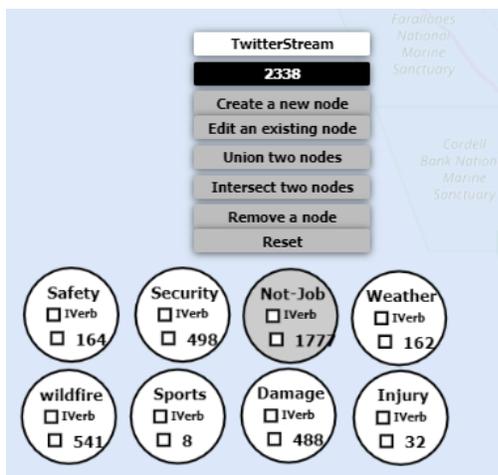

SMART Classifiers allow users to observe, search, and classify tweets by user-defined keywords (using inclusion and exclusion of such keywords). When a user creates a classifier, they are required to supply a name (e.g., "Security") and provide a set of keywords. Enabling a classifier will identify and display messages containing one or more of the classifier's keywords. The user can also union and intersect classifiers for further analysis.

### 2) Content Lens

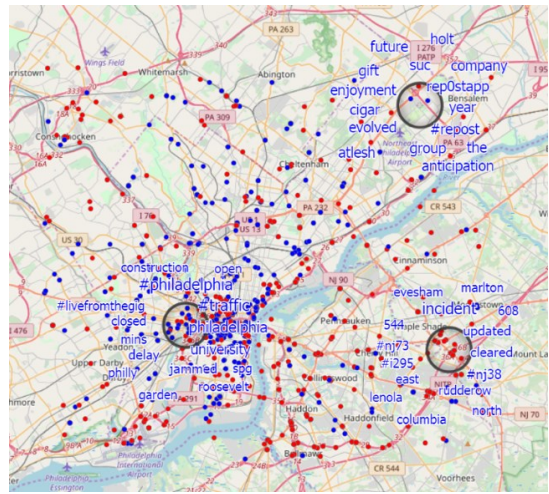

The content lens feature allows users to place lenses with word-cloud visualization over specific geographical locations. The area immediately surrounding the lens will be populated with the most frequently used keywords within the lens location, using integrated topic modeling for the geographic area under the lens.

### 3) Cluster Lens

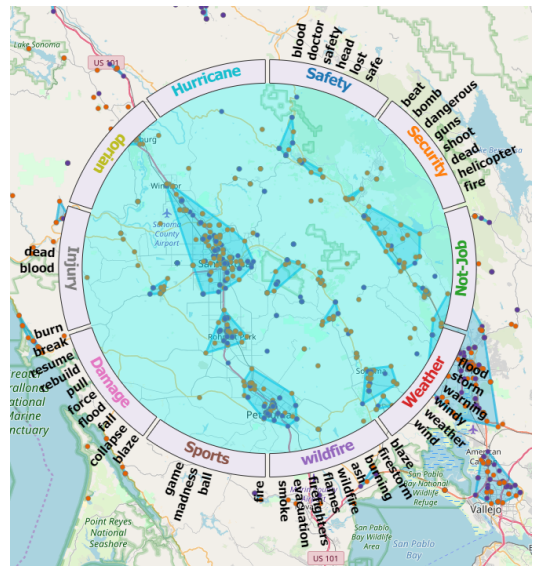

The cluster lens filters message clusters within the lens and visualizes the keywords related to the classifiers in a radial layout. The user can move the underlying map to investigate regions of interest while the position of the lens remains fixed on the screen.

*4) Tweet Tooltip*

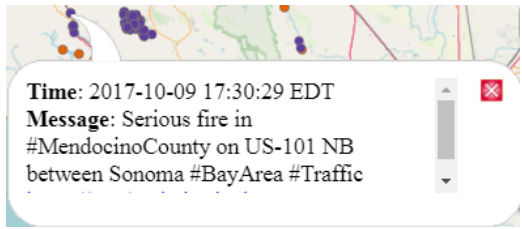

The tweet tooltip provides users with the ability to view a message's content directly from the map by clicking on the message's symbol.

*5) Heatmap*

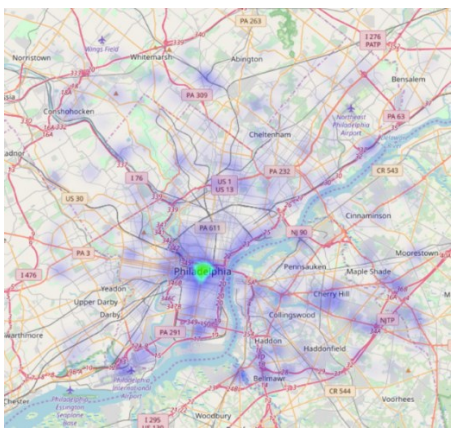

SMART's heatmap provides an overview of the spatial distribution of the tweets.

*6) Spatial Filtering*
   *a) Geo-fencing*

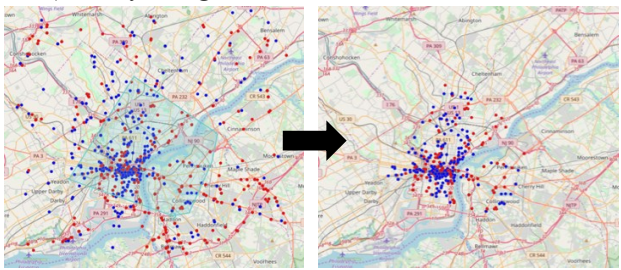

The geo-fencing tool allows users to define a geographical bound and remove messages outside of the bound from the visualizations.

*b) Region Selection*

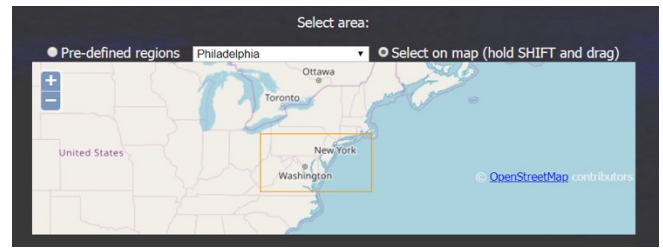

SMART's region selection feature is provided on system start-up. Before loading the main system interface, users can select a specific region on the map to stream messages from during system use.

*7) Temporal View (Time Bar Graph)*

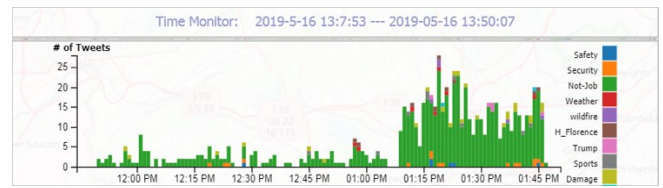

The temporal view displays the number of tweets in different classifiers over time. It is aggregated at the minute level and reflects the evolution over the last two hours.

*8) LDA Topic Modeling View*

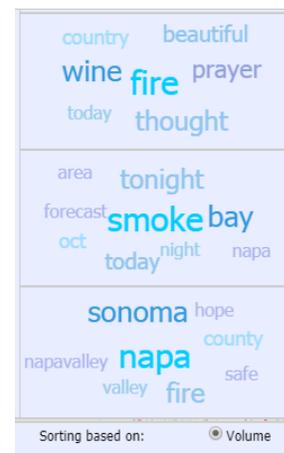

The LDA topic modeling view uses Latent Dirichlet Allocation to extract trending topics and associated keywords among the messages. Users can click on a colored keyword to locate messages containing it.